# SCOR: A Framework for Responsible AI Innovation in Digital Ecosystems

Mohammad Saleh Torkestani
*faculty of environment science and economy*
*University of Exeter*
Exeter, United Kingdom
m.torkestani@exeter.ac.uk

Taha Mansouri
*School of Science, Engineering and Environment*
*University of Salford*
Manchester, United Kingdom
t.mansouri@salford.ac.uk

**Abstract**

AI-driven digital ecosystems span diverse stakeholders including technology firms, regulators, accelerators and civil society, yet often lack cohesive ethical governance. This paper proposes a four-pillar framework (SCOR) to embed accountability, fairness, and inclusivity across such multi-actor networks. Leveraging a design science approach, we develop a Shared Ethical Charter(S), structured Co-Design and Stakeholder Engagement protocols(C), a system of Continuous Oversight and Learning(O), and Adaptive Regulatory Alignment strategies(R). Each component includes practical guidance, from "lite" modules for resource-constrained start-ups to in-depth auditing systems for larger consortia. Through illustrative vignettes in healthcare, finance, and smart city contexts, we demonstrate how the framework can harmonize organizational culture, leadership incentives, and cross-jurisdictional compliance. Our mixed-method KPI design further ensures that quantitative targets are complemented by qualitative assessments of user trust and cultural change. By uniting ethical principles with scalable operational structures, this paper offers a replicable pathway toward responsible AI innovation in complex digital ecosystems.

## 1 Introduction

Digital ecosystems, marked by multi-stakeholder collaboration across technology firms, regulators, investors, accelerators and civil society, are increasingly propelled by artificial intelligence (AI) (Iansiti & Levien, 2004; Jacobides, Cennamo & Gawer, 2018). While AI accelerates innovation and value creation, its pervasive impact on societal structures (healthcare, finance, public services, and more) raises pressing ethical, regulatory, and governance challenges (Cath, 2018; Leslie, 2019). Researchers, policymakers, and industry leaders increasingly recognize that responsible AI development requires frameworks addressing fairness, accountability, transparency, and inclusivity across organizational boundaries (Floridi, 2019; London, 2024).

However, existing high-level AI ethics guidelines, whether from academia (Jobin, Ienca & Vayena, 2019) or industry associations (IEEE, OECD, WEF), often stop short of detailing how multi-stakeholder ecosystems can operationalize these principles systematically (Minkkinen, Zimmer, et al., 2022). In contexts where data flows, product roadmaps, and regulatory obligations span multiple jurisdictions and corporate structures, conventional single-organization compliance models prove insufficient (Moore, 1993; Gawer & Cusumano, 2014). Furthermore, the role of organizational culture and leadership incentives cannot be understated: if senior management remains focused on short-term growth metrics, ethical oversight may be sidelined (London, 2024). Simultaneously, resource constraints can impede smaller firms' ability to invest in robust auditing or inclusive co-design processes (Shams, Zowghi & Bano, 2023).

Against this backdrop, the present study proposes and refines a four-pillar ethical compliance framework tailored for AI-driven ecosystems. Building on design science (Hevner et al., 2004; Peffers et al., 2007), we synthesize scholarship on AI ethics (Leslie, 2019), ecosystem governance (Jacobides et al., 2018), diversity and inclusion (Zowghi & da Rimini, 2022; Shams, Zowghi & Bano, 2023), and organizational behavior (London, 2024). Our framework underscores Shared Ethical Charter, Co-Design & Stakeholder Engagement, Oversight, and Regulatory Alignment.

Each component focuses on aligning multiple stakeholders around ethical principles, practical governance structures, and data-driven metrics that encourage continuous improvement. In addition to describing these components, the paper includes illustrative vignettes to show how they might function in healthcare, finance, or smart city contexts. To address issues of scalability, particularly relevant for start-ups and SMEs, scaled-down ("lite") governance modules and resource-sharing initiatives are also considered.

The rest of the paper proceeds as follows: Literature Review explores key themes in AI ethics, ecosystem theory, leadership incentives, and co-design. Methodology discusses our design science approach, including how we integrated multiple data sources. Findings & Analysis presents the framework's components, enhanced with hypothetical examples, best practices, and recommended metrics. Discussion critically examines real-world feasibility, resource constraints, stakeholder alignment, and how to operationalize the framework for SMEs and large consortia alike. Conclusion synthesizes contributions, outlines future research (including pilot designs), and highlights our framework's distinctiveness vis-à-vis existing guidelines.

By integrating conceptual rigor with practical guidance, this paper aims to equip ecosystem actors (large technology firms, accelerators, smaller enterprises, regulators, NGOs) with a scalable yet flexible roadmap for responsible AI in complex digital environments.

# 2 Literature Review

## 2.1 Governance in AI-Driven Digital Ecosystems

### 2.1.1 Evolving Ecosystem Theory

The notion of digital ecosystems builds on Moore's (1993) concept of business ecosystems, recognizing that innovation often emerges from interconnected actors (producers, partners, users) co-creating value (Iansiti & Levien, 2004; Adner, 2017). In these ecosystems, AI now serves as a "keystone technology," enabling sophisticated collaborations but also creating multi-actor dependencies (Jacobides, Cennamo & Gawer, 2018; Minkkinen, Zimmer et al., 2022). Consequently, AI governance cannot rely solely on organizational-level compliance. Rather, ecosystem-level frameworks (encompassing technical, legal, ethical, and social dimensions) are necessary (Cath, 2018).

### 2.1.2 Accountability, Leadership, and Organizational Culture

A recurring theme in AI governance literature is the role of senior leadership in cultivating ethical culture (London, 2024). Traditional compliance hinges on top-down policies, but it may fail if organizational incentives prioritize market speed overdue diligence (Floridi, 2019). Scholars thus underscore the need for ethics to pervade culture, ensuring that mid-level managers and technical staff have the autonomy and support to raise ethical concerns (Owen, Macnaghten & Stilgoe, 2012). This implies incentive realignment, for example, tying executive KPIs to ethical outcomes, building internal training programs, and forming cross-disciplinary ethics boards with genuine veto power (Leslie, 2019; London, 2024).

Yet, accountability structures easily falter in ecosystems where responsibilities diffuse across multiple stakeholders. Multi-stakeholder councils and public-private consortia can mitigate the fragmentation by jointly defining shared principles and oversight bodies (WEF, 2022). This approach fosters collective accountability: each participant commits to transparency, fairness, and inclusivity beyond corporate boundaries (Cath, 2018).

## 2.2 Diversity, Inclusion, and Stakeholder Co-Design

### 2.2.1 The Current State of DEI in AI

A growing body of work emphasizes diversity and inclusion (D&I) as critical to AI's ethical and social legitimacy (Zowghi & da Rimini, 2022; Shams et al., 2023). For instance, Shams et al. (2023) identify at least 55 unique challenges (from gender bias in training data to lack of inclusive user interfaces) that hamper AI's fairness and trustworthiness. The interplay between algorithmic bias and societal discrimination can magnify historical inequities (Eubanks, 2019; Benjamin, 2019). Proposed solutions vary, including community co-design, which engages marginalized groups in model development (Shams et al., 2023); fostering diverse AI teams through more inclusive hiring and training practices (Omidvar et al., 2021); and ensuring data sets reflect demographic and cultural heterogeneity to enhance inclusivity (WEF, 2022).

### 2.2.2 Imperatives for DEI in AI

Extensive work highlights that bias can be "designed in" or amplified by AI (O'Neil, 2017; Eubanks, 2019). Diversity and inclusion (D&I) thus represent an ethical and strategic priority.

Underrepresented users, whether due to gender, race, language, or socioeconomic status, risk adverse treatment if their data are missing or misrepresented (Shams, Zowghi & Bano, 2023). However, technical fixes alone (e.g., de-biasing algorithms) are insufficient unless development teams themselves reflect diverse backgrounds and incorporate the lived experiences of marginalized communities (Zowghi & da Rimini, 2022).

### 2.2.3 Stakeholder Co-Design and Power Dynamics

Co-design methodologies (systematically involving end-users, community advocates, and domain experts) encourage direct participation from impacted communities throughout the AI lifecycle (Owen et al., 2012). This approach can enhance contextual awareness of potential harms and opportunities, improve trust by ensuring transparency and user empowerment, and expose power differentials (e.g., corporate capture, profit motives) early (Ansell & Gash, 2008).

However, effective co-design requires protocols for participant selection, balanced facilitation, systematic feedback loops, and conflict resolution (Shams et al., 2023). Without such rigor, co-design can become tokenistic or overshadowed by dominant interests. This underscores the need for clear governance structures and robust cultural support from leadership (WEF, 2022).

## 2.3 3. Responsible Innovation and the Socio-Technical Lens

### 2.3.1 Socio-Technical Complexity

AI systems are socio-technical constructs: they intertwine algorithmic processes with human values, organizational structures, and cultural contexts (Floridi & Taddeo, 2018). Scholars highlight the danger of focusing on technical fixes (e.g., better algorithms) without addressing social or organizational root causes, like prejudicial data generation or homogenous decision-making teams (Eubanks, 2019). Effective governance, therefore, requires holistic risk assessments that capture organizational readiness, stakeholder trust, and technical vulnerabilities (Floridi, 2019; Leslie, 2019); stakeholder involvement, particularly from underrepresented groups, throughout design phases (Zowghi & da Rimini, 2022; Shams et al., 2023); and iterative re-alignment of technical systems with evolving ethical norms (Cath, 2018).

### 2.3.2 Actionable Frameworks for Ecosystem-Level Compliance

A key gap is how to operationalize these socio-technical perspectives at scale. Building on responsible innovation scholarship (Owen, Macnaghten & Stilgoe, 2012), ecosystem-level solutions may combine design science through iterative artifact creation and evaluation, bridging ethical principles with pragmatic governance tools (Hevner et al., 2004; Mäntymäki et al., 2022); co-regulatory or hybrid governance via partnerships among governments, corporations, and NGOs to develop shared standards, possibly through technical documentation or auditing frameworks (Cath, 2018; WEF, 2022); and transparency infrastructure by mandating machine-readable ethics labels, algorithmic impact assessments, or public registers of AI models (Leslie, 2019).

While these emergent frameworks remain heterogeneous, they consistently stress multi-stakeholder collaboration, robust accountability mechanisms, and continuous learning. The next sections build on these insights by proposing a design science-driven methodology to craft and refine a replicable ecosystem-level ethical compliance framework, aiming to embed fairness, inclusivity, and accountability in AI-driven digital ecosystems.

## 2.4 Existing Frameworks: Similarities and Gaps

### 2.4.1 Overview of Notable Guidelines

Several organizations have proposed guidelines for ethical AI: Leslie (2019) outlines a process-based governance approach for the public sector, emphasizing interpretability, accountability, and data stewardship. IEEE (2019) suggests Ethically Aligned Design, focusing on human-centric AI. OECD has AI principles codified in broad policy statements, while the WEF publishes toolkits (2022) for inclusive AI. Mäntymäki et al. (2022) propose the "hourglass model," distinguishing governance layers at environment, organization, and system levels.

Though robust, these frameworks often focus either on organizational or public-sector contexts. Further, many stop short of detailing how multi-stakeholder co-design might be orchestrated or how to measure progress across supply chains with mixed resource capacities (Lin, 2020; Minkkinen et al., 2022).

### 2.4.2 Differentiating Our Proposed Framework

Our approach builds on previous work but extends it in several key areas by emphasizing ecosystem-level depth through continuous collaboration among corporations, SMEs, regulators, and community groups while incorporating adaptive regulatory alignment. It acknowledges that no single entity can ensure holistic compliance, advocating for multi-actor synergy. The framework follows a structured four-pillar design consisting of a Shared Ethical Charter, Co-Design Mechanisms, Continuous Oversight, and Adaptive Alignment, each with KPIs and recommended sub-processes to bridge principle and practice. With a strong focus on practical implementation, it offers concrete structures, including step-by-step protocols, vignettes, and potential "lite" versions for SMEs to enhance accessibility. Measurement clarity is ensured through specific quantitative and qualitative KPIs to prevent tokenistic box-ticking. Additionally, the framework is pilot-ready, outlining empirical validation through pilot testing or action research. By providing expanded KPI detail, concrete usage scenarios, and guidance for smaller players, we aim to develop a holistic framework that practitioners can adapt across multiple jurisdictions and varying resource levels (Cath, 2018; WEF, 2022).

## 2.5 Empirical Validation and Pilot Designs

Empirical validation and pilot designs are crucial for refining conceptual frameworks that often remain unvalidated in large-scale practice (Stahl et al., 2017, Shams et al., 2023). Action research or pilot implementations, such as in smart city infrastructure or AI-based recruitment, enable iterative refinement by assessing pre-existing governance norms, data flows, and stakeholder engagement, followed by implementing each framework component for a designated period. Measurement involves tracking adoption rates, fairness outcomes, user satisfaction, and cross-jurisdiction consistency, with a continuous learning loop to adjust the framework based on findings and re-measure for improvements. Future research questions include examining cultural variation in interpreting co-design and Charter obligations across different legal contexts (EU, US, Asia), assessing how the framework scales in small local consortia versus large multinational alliances, and understanding long-term governance dynamics, particularly whether oversight committees retain influence over multiple product cycles or if "ethics fatigue" emerges. Addressing these questions would enhance the framework's transferability and credibility across diverse digital ecosystems.

## 3   Methodology

### 3.1 Design Science Approach

Following Hevner et al. (2004), Peffers et al. (2007), and Gregor and Hevner (2013) we employed a Design Science (DS) methodology to construct, illustrate, and refine our four-pillar ethical compliance framework. DS is well-suited for socio-technical challenges, blending theoretical principles with practical artifacts. We began with problem identification, observing a gap between high-level AI ethics ideals and ecosystem-level actionable frameworks (Jobin et al., 2019; Leslie, 2019). Our objective was to create a replicable model for multi-stakeholder alignment around ethics, accountability, and inclusivity. For design and development, we drew on academic works (Cath, 2018; Minkkinen et al., 2022) and policy documents (EU AI Act, OECD guidelines) to draft the four components. In the demonstration phase, we integrated vignettes and hypothetical examples to illustrate how each component might function in healthcare and finance settings, responding to calls for more practical detail. Although we lack direct empirical pilot data, we performed theoretical cross-checks by comparing our approach to IEEE, WEF, and Leslie's frameworks, and we propose a future pilot design for further validation. Finally, this paper communicates the final artifact and outlines next steps, including pilot testing and potential extensions to SMEs.

### 3.2 Data Sources

The data sources for this study include academic literature, consisting of peer-reviewed articles on AI ethics, ecosystem theory, co-design, and leadership incentives; policy and industry materials, such as regulatory texts (European Commission, 2021), cross-sectoral white papers (OECD, WEF), and big-tech ethical guidelines; and expert discussions, incorporating insights from AI governance forums, domain-specific conferences, and multi-stakeholder roundtables.

### 3.3 Potential Pilot Study Outline

To move beyond conceptual design, a potential pilot study could be structured by selecting a sector with clear ethical stakes, such as AI-driven recruitment or healthcare diagnostics, and multi-actor involvement. The framework would be introduced over a 6-12 month period, involving the formation of a Shared Ethical Charter, co-design sessions, regulatory compliance mapping, and periodic audits. Data collection would track both quantitative indicators, such as "Implementation Ratio" and "Audit Coverage," and qualitative measures, including user trust and staff perceptions. Analysis would compare baseline and post-implementation results, documenting improvements, tensions, or emergent best practices. The refinement phase would use these insights to iterate on the framework's protocols, KPIs, or governance structures. Ultimately, such a pilot would generate empirical evidence on the framework's effectiveness and scalability, supporting its broader adoption across digital ecosystems.

## 3.4 Limitations of the Method

While DS offers a robust approach for bridging theory and practice, real-world complexities (varying resource levels, leadership commitments, legal contexts) necessitate contextual adaptation. No single model can fully account for cross-cultural or sector-specific nuances (Lin, 2020). Consequently, future expansions may refine specific guidelines for domains (e.g., AI in public health vs. finance) or legal jurisdictions.

# 4 Findings & Analysis

This section details the four-pillar framework, highlighting operational details, illustrative vignettes, and scalability strategies.

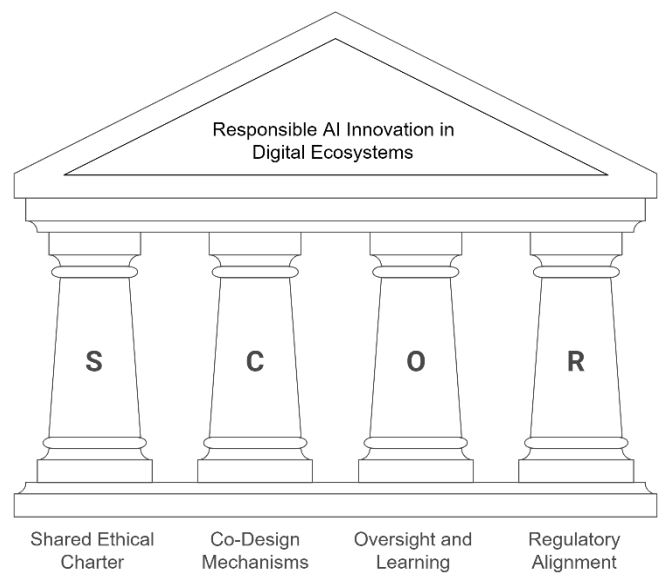

Figure 1: Proposed framework (authors)

## 4.1 The Four-pillar Ethical Compliance Framework (SCOR)

SCOR stands for Shared Ethical Charter, Co-Design & Stakeholder Engagement, Oversight, and Regulatory Alignment, key pillars for governing AI-driven digital ecosystems. To ensure these principles translate into practice, each pillar is supported by measurable Key Performance Indicators (KPIs). The KPIs balance quantitative and qualitative metrics and are designed for business-level monitoring by an independent ecosystem audit body. They go beyond generic counts to focus on meaningful compliance and performance, supporting both internal self-assessment and external auditing.

### 4.1.1 Shared Ethical Charter (S)

The Shared Ethical Charter serves as a binding set of ethical commitments (fairness, accountability, transparency, and inclusivity) agreed upon by all ecosystem participants (Floridi, 2019; Leslie, 2019). The Charter should establish baseline requirements, such as prohibiting the deployment of discriminatory AI, ensuring user recourse, and mandating routine audits. It must also define enforcement mechanisms, including publicly documented processes for flagging violations and designated committees or boards with the authority to impose sanctions or corrective actions. Signatory responsibilities involve commitments to training, data governance compliance, and adherence to co-design protocols. The following KPIs ensure the charter's principles are widely adopted, internalized, and enforced across organizations:

- **Charter Adoption Rate:** Percentage of partner organizations that have formally endorsed the shared ethical charter and integrated its principles into internal policies. A high adoption rate (target ~100%) indicates ecosystem-wide commitment, moving ethical charters from mere statements to actionable commitments (Pistilli, et al. 2023). This can be verified by each organization's documented approval of the charter and evidence of its principles in corporate guidelines.

- **Ethics Training & Awareness:** Proportion of employees and relevant stakeholders trained on the charter's principles (e.g. completed annual ethics training). Measured via training completion rates and assessment scores, this KPI ensures that the charter is not just signed at the top but understood at all levels. High rates (e.g. >90% staff trained) and strong scores indicate that ethical values are being disseminated and absorbed across the ecosystem.

- **Ethical Decision Review Coverage:** Percentage of high-impact AI projects or decisions that undergo a formal ethics review for charter alignment before deployment. For example, an ethics committee or review board should vet whether a new AI system adheres to charter principles (fairness, transparency, etc.). A target of 100% for designated high-risk or critical AI initiatives ensures no major system goes live without scrutiny for ethical compliance.

- **Charter Compliance Incidents Resolved:** Number of reported breaches of the ethical charter and the percentage of those incidents that are addressed through formal remediation or sanctions. This KPI tracks how well the ecosystem enforces the charter, e.g. if an audit finds a violation of the agreed principles, was it corrected promptly and were there consequences? (For instance, in traditional business ethics, if employees violate a company's ethical charter, it triggers internal sanctions by the company or its ethics committee (Pistilli, et al. 2023).) A low incidence of breaches, coupled with 100% remediation of identified issues, would reflect strong compliance.

### 4.1.2 Co-Design and Stakeholder Engagement Mechanisms (C)

The Co-Design and Stakeholder Engagement Mechanisms ensure regular, structured engagements that bring together domain specialists, underrepresented community members,

NGO advocates, and regulators (Owen et al., 2012). Co-design fosters diverse input, minimizing blind spots related to bias or cultural insensitivity; trust, as participants gain visibility into system design while developers better understand user needs; and conflict resolution, allowing early negotiation of tensions such as profit versus fairness. Operationally, roundtable cadence should be monthly or quarterly, with rotating representatives to maintain fresh perspectives. Facilitation requires a neutral ethics officer or external mediator to prevent dominant voices from overshadowing discussions. Documentation should include meeting minutes summarizing proposed changes, the rationale for acceptance or rejection, and follow-up actions. KPIs include the Stakeholder Representation Index, a weighted measure capturing demographic and expertise diversity; the User Recommendation Implementation Ratio, reflecting the percentage of co-design suggestions adopted versus deferred; and Engagement Quality, assessed through post-session feedback forms evaluating participants' sense of inclusion, clarity, and influence over decisions. In a finance application, a cross-border fintech alliance deploying credit-risk AI could hold bimonthly roundtables with local consumer advocacy groups, data scientists, and government representatives. Each session would address model fairness outcomes, ensuring that concerns such as credit invisibility among marginalized communities are integrated into system updates. Summaries of these discussions would be publicly accessible on a consortium website. KPIs below measure meaningful engagement quality rather than just headcounts:

- **Stakeholder Inclusion in AI Projects:** Percentage of AI projects that include multi-stakeholder consultation or co-design sessions during key development phases. This can be measured by checking project records for involvement of external stakeholders (e.g. user representatives, ethicists, affected community members) in design workshops or requirements reviews. A high inclusion rate signifies that co-design is standard practice, not an exception (Leslie, 2019). For example, teams should seek stakeholder input on the acceptability of project plans, engaging in "collaborative deliberation" beyond the core team about ethical impacts of the design (Leslie, 2019).

- **Diversity of Stakeholder Representation:** An index or score assessing the diversity of stakeholders engaged (across gender, ethnicity, domain expertise, etc.) in governance forums or design processes. This qualitative-but-measurable KPI ensures that not just a token group of stakeholders is consulted, but a broad spectrum reflecting those impacted. Audit bodies can review stakeholder rosters for representation of vulnerable or under-represented groups. The goal is to cover all key stakeholder categories so that AI solutions meet the needs of all and do not inadvertently harm absent voices (WEF, 2022).

- **Stakeholder Feedback Implementation Rate:** Measures the extent to which input from stakeholders is acted upon. For instance, of the recommendations or concerns raised by stakeholders during co-design, what percentage are either implemented or formally addressed in the final design? A high implementation rate indicates genuine empowerment of stakeholder voices (not just perfunctory meetings). Documentation from design reviews can be audited to see how stakeholder recommendations influenced outcomes. This KPI moves beyond counting meetings to gauging the impact of engagement.

- **Stakeholder Impact Assessment (SIA) Completion:** Percentage of major AI initiatives that undergo a formal Stakeholder Impact Assessment before launch. An SIA systematically evaluates a project's societal and ethical impact on different stakeholder groups (Leslie, 2019). A target of 100% for high-risk or public-facing AI systems means every such project is vetted for how it might affect communities and stakeholders, with findings used to improve design. This aligns with best practices to "come together to evaluate the social impact and sustainability" of AI projects through SIA, giving due consideration to those affected (Leslie, 2019).

- **Stakeholder Trust/Satisfaction Score:** A qualitative KPI obtained via periodic surveys or feedback from external stakeholders about their trust in and satisfaction with the AI systems or the engagement process. For example, community representatives or end-users could rate the transparency, fairness, or responsiveness of the AI provider. While subjective, this is quantified through scores (e.g. on a 5-point scale or Net Promoter Score style). Improvement in this metric over time would indicate that co-design and engagement efforts are building public trust in the AI ecosystem, addressing the risk that poor inclusion "erodes trust in communities" (WEF, 2022). An ecosystem audit body might anonymously survey stakeholders of multiple member organizations to derive an overall trust index.

### 4.1.3 Oversight and Learning (O)

The Ongoing Audits and Governance Mechanisms framework ensures continuous oversight, preventing ethical stagnation or "compliance drift" (Leslie, 2019; Mäntymäki et al., 2022). Key components include periodic audits, scheduled or random, to evaluate model performance, data handling, and bias incidence; an Ecosystem Governance Board, composed of industry representatives, regulators, NGOs, and user communities, responsible for reviewing audit findings, enforcing remedial actions, and updating best practices; and a knowledge repository, a central hub compiling near-miss incidents, success stories, and evolving guidelines to encourage collective learning and transparency. Below KPIs aim to strengthen oversight mechanisms at the ecosystem level:

- **Internal AI Governance Committee Presence:** Percentage of organizations in the ecosystem that have an internal AI ethics or governance committee (or designated AI ethics officer) in place. This binary KPI (yes/no per organization) is rolled up to ensure, for instance, "100% of member companies have established a cross-functional AI oversight committee." Such committees should meet regularly to review AI use cases and risks. Their existence and activity can be verified through organizational charts and meeting minutes. This reflects emerging best practice: companies like OneTrust formed dedicated AI governance committees to ensure AI use conforms to responsible AI principles and regulations.

- **Independent Audit Coverage:** Proportion of AI systems (or projects) that undergo independent external audit or assessment for ethical compliance each year. This quantitative KPI monitors the reach of third-party oversight, e.g. an annual ethics audit of all "high-risk" AI applications. A higher coverage (aiming for 100% of critical systems audited) signifies robust accountability. It addresses the current gap where only ~46% of organizations have independent ethical audits for AI. Improvement in this metric over time means more of the AI portfolio is being scrutinized by external experts, aligning with accountability standards.

- **Audit Trail and Documentation Quality:** An indicator of how well AI projects maintain documentation to enable end-to-end oversight. This could be measured as the percentage of AI projects that have complete "audit logs" and documentation of data provenance, model development, and decision rationale, as checked by the audit body. According to best practices, an AI project must be "accessible for audit, oversight, and review," which requires builders to keep detailed records of data and processes (Leslie, 2019). This KPI ensures organizations are "audit-ready", e.g. scoring each project's documentation completeness, with targets to close any gaps.

- **Oversight Recommendations Implementation:** Measures the responsiveness of management to oversight findings. For instance, what percentage of the recommendations or corrective actions identified by the ecosystem's audit body (or internal oversight committees) are implemented within an agreed timeframe? A high percentage (or short average closure time for issues) means oversight has real teeth and leads to tangible improvements. This KPI can be tracked by following each audit report's action items to completion, ensuring that oversight is not just nominal but enforceable.

- **Incident Reporting & Remediation Rate:** Number of AI ethics or compliance incidents reported (e.g. identified cases of bias, privacy violation, or system failure with harm) and the percentage of those incidents that are resolved through the oversight process. This encourages transparent reporting of problems and swift remediation. The audit body can require each organization to log incidents and report how they were addressed. A low number of unresolved incidents (or none at all) indicates effective oversight and risk management. Conversely, a spike in incidents might trigger deeper investigation by the ecosystem overseers.

### 4.1.4 Regulatory Alignment (R)

The Adaptive Regulatory Alignment mechanism ensures that ecosystems anticipate and comply with evolving regulations, such as the EU AI Act, national privacy laws, or industry-specific guidelines (Cath, 2018; Lin, 2020). Key elements include regulatory sandboxes, which are government-supervised pilot programs that enable controlled trials of new AI solutions; horizon scanning committees, composed of multi-actor teams that monitor legislative changes and guidance from standard-setting bodies like IEEE and OECD; and global/local interoperability, ensuring coordination in cross-border data usage or algorithm deployments, potentially adopting the strictest region's norms as a baseline for safety. These KPIs keep organizations in sync with legal requirements:

- **High-Risk AI Compliance Rate:** Percentage of AI systems categorized as "high-risk" (under frameworks like the EU AI Act) that fully meet all regulatory requirements before deployment. This includes completing mandatory risk assessments, documentation, and obtaining any required certifications. The KPI could be measured via compliance checklists or certifications obtained. The goal is 100% compliance for high-risk AI, since such systems will face strict obligations (e.g. transparency, human oversight) once regulations take effect. Upcoming laws take a risk-based approach, high-risk uses (like AI in critical infrastructure or hiring) "require a higher degree of legal compliance" (Pistilli, et al. 2023), so this KPI tracks conformance in those critical cases.

- **Regulatory Audit/Certification Pass Rate:** Proportion of AI initiatives or organizations that pass external regulatory audits or obtain certifications each year. For example, if a data protection authority or an AI regulator conducts an audit, did the organization pass without major findings? Or, how many AI systems achieved compliance seals (such as a CE marking under the EU AI Act's conformity assessment)? A high pass rate demonstrates that the ecosystem not only internally monitors ethics, but also consistently meets the bar set by independent regulators.

- **Timely Regulatory Update Implementation:** Measures how quickly organizations adapt to new or updated AI regulations or guidelines. For instance, when a new law or standard is introduced, how many months do organizations take on average to update their processes or systems accordingly? This can be tracked by the audit body through compliance reports. Short turnaround times (e.g. implementing required changes within a few months of a new rule) indicate agility and commitment to alignment. This KPI prevents lagging compliance, ensuring companies aren't caught unprepared by audits or legal changes.

- **Mandatory Impact Assessment Completion:** Percentage of AI systems that undergo legally required impact assessments (such as Algorithmic Impact Assessments or Data Protection Impact Assessments) in jurisdictions where they are required. For example, if a law like the proposed US Algorithmic Accountability Act mandates impact assessments for AI systems using personal data, the KPI tracks compliance with that mandate. A near-100% rate indicates organizations are diligently performing these analyses. This aligns with emerging regulations: the Algorithmic Accountability Act would require organizations to conduct impact assessments of automated decision systems handling personal information ().

- **Regulatory Incident Rate:** Number of regulatory violations, fines, or legal sanctions related to AI incurred by ecosystem members, tracked annually. The ideal target is zero. If any occur, the audit body would examine them, but tracking this KPI over time helps the ecosystem identify problem areas. A consistently zero or very low incident rate means that alignment mechanisms (policies, checks, and audits) are effectively preventing non-compliance. If incidents do occur, this KPI coupled with oversight KPIs ensures there is a feedback loop to fix the underlying compliance gaps.

## 4.2 SCOR Model Implementation Vignettes

The SCOR framework can be applied across industries to strengthen ethical AI governance. Below, each pillar is illustrated with a real-world scenario, highlighting key stakeholders and demonstrating how SCOR leads to ethical, inclusive, and accountable AI governance in practice.

### 4.2.1 Shared Ethical Charter (S)

**Scenario:** In the healthcare sector, a national health service partners with technology companies to establish a shared ethical charter for AI in patient care. For example, the UK's NHS created an AI code of conduct with input from industry, academics, and patient advocates (GOV.UK, 2019). This charter outlines common principles, like patient safety, transparency, fairness, and data privacy, that all AI systems must uphold before they can be deployed in hospitals. By agreeing on these ethical guidelines upfront, every stakeholder has a clear understanding of the "rules of the road" for AI solutions.

- **Key Stakeholders:** Government health agencies (setting the charter), healthcare providers and clinicians, AI developers, patient advocacy groups, and regulators. All parties collectively define and commit to the ethical principles.

- **Impact on AI Governance:** The shared charter ensures alignment of values across organizations, making AI deployments inherently safer and more inclusive. It builds public trust (patients know that AI diagnostics meet strict ethical standards) and establishes accountability, if an AI system violates the charter (e.g. by compromising privacy or fairness), stakeholders are obligated to address it or halt its use. This pillar makes ethical principles a tangible part of AI development, creating a foundation for responsible AI governance across the industry.

### 4.2.2 Co-Design and Stakeholder Engagement Mechanisms (C)

**Scenario:** A city government is deploying an AI-driven traffic management system as part of a smart city initiative. Instead of a top-down implementation, the city uses co-design and stakeholder engagement: it invites local residents, urban planners, public transit operators, and disability advocacy groups to help design and refine the system. Through community workshops and feedback sessions, these stakeholders influence how the AI will balance goals (e.g. reducing congestion and preserving walkability in neighborhoods). This participatory design process ensures the AI system reflects community values and concerns from the start (Saxena et al., 2025).

- **Key Stakeholders:** City officials and planners, the technology vendor's design team, everyday citizens (commuters, residents of affected neighborhoods), civil society groups (accessibility and environmental advocates), and even traffic police. Each brings unique insights, from lived experiences of traffic issues to technical and ethical considerations.

- **Impact on AI Governance:** By engaging stakeholders in co-design, the AI solution becomes more inclusive and context-aware. It is tuned to serve all community members

(for instance, avoiding bias toward wealthy districts or car owners) and addresses potential harms raised by the public before they occur. Ongoing stakeholder involvement (e.g. a citizen advisory panel) provides accountability: the system's outcomes (like changes in traffic patterns) are monitored and discussed with the community, making the AI's operators answerable to those it affects. This collaborative approach builds trust in the technology, residents are more likely to accept and support an AI system they had a hand in shaping, illustrating how participatory governance leads to more ethical AI use.

### 4.2.3 Oversight and Learning (O)

**Scenario:** A global software company implements a robust AI oversight structure to govern its many AI products. For example, enterprise software maker SAP has both an external AI ethics advisory board and an internal AI ethics committee working in tandem (Yokoi and Wade, 2023). In practice, the company assembles a panel of external experts (academics, ethicists, industry experts) alongside senior internal stakeholders (AI lead engineers, risk officers, legal counsel). This oversight board regularly reviews high-risk AI projects, such as a new AI-driven hiring tool or an algorithm that optimizes factory workflows and evaluates them against ethical standards and the company's own AI principles. They have the authority to suggest improvements or even veto deployments if an AI system poses undue ethical risks.

- **Key Stakeholders:** Internal leadership and AI development teams, external domain experts and ethicists, employee representatives, and sometimes regulator observers. The mix ensures a 360-degree view, technical feasibility, ethical implications, legal compliance, and public perception are all considered.

- **Impact on AI Governance:** This pillar creates a continuous accountability mechanism inside the organization. Oversight means AI systems are not "fire and forget", they are continuously monitored and audited. Issues like bias, safety incidents, or misuse are caught early by the oversight committee and addressed transparently. The presence of external advisors also adds independent scrutiny, bolstering trust with regulators and the public that the company isn't "marking its own homework." Overall, an oversight board enforces the AI's alignment with the shared ethical charter (Pillar 1) and societal norms, ensuring the company remains **accountable** for its AI's behavior and maintains high ethical standards even as it innovates.

### 4.2.4 Regulatory Alignment (R)

**Scenario:** A multinational bank is rolling out an AI-powered credit scoring system across several countries. To do so responsibly, it emphasizes regulatory alignment at every step. The bank's AI team works closely with legal and compliance departments to ensure the system meets all relevant laws and guidelines in different jurisdictions. For example, in the US they adhere to fair lending regulations and EEOC guidelines, and even proactively comply with local rules like New York City's requirement for bias audits in automated hiring tools (Smith, 2023) (a similar principle applied to credit decisions). In the EU, they anticipate the upcoming AI Act by categorizing their credit AI as high-risk and implementing the Act's expected transparency and oversight measures (Smith, 2023). The bank even participates in regulatory

"sandboxes", collaborative programs with central banks and regulators, to test its AI in a supervised environment and refine it according to feedback.

- **Key Stakeholders:** The financial institution's AI developers and compliance officers, banking regulators and data protection authorities in each region, external auditors or certification bodies, and consumers (borrowers) who are protected by these regulations.

- **Impact on AI Governance:** Aligning with regulations ensures the AI system is accountable under the law and respects citizens' rights. This pillar forces clarity and rigor, features like explainability and bias mitigation are built in to satisfy regulatory scrutiny, which in turn makes the AI more transparent and fairer for consumers. By meeting strict standards (e.g. undergoing independent audits to prove no discrimination (Smith, 2023), the bank's AI earns greater trust from both regulators and the public. Regulatory alignment also promotes inclusivity: laws often require consideration for underrepresented or vulnerable groups, so the AI's design accounts for these from the outset. In sum, this pillar operationalizes ethical AI principles through compliance, the AI not only follows the company's values, but also upholds societal rules and norms across regions, leading to broadly trustworthy and responsible AI deployment.

## 4.3 Feasibility, Scalability, and Resource Sensitivity

### 4.3.1 Scaled-Down Modules for SMEs

Smaller firms often face resource limitations that hinder their ability to conduct extensive audits or participate continuously in co-design roundtables (Shams et al., 2023). Possible solutions include Shared Ethics Boards, where SMEs pool funds to hire experts who provide cross-ecosystem oversight; Tiered Charter Compliance, focusing initially on two to three high-risk areas such as privacy and bias, with an expanding scope as the SME grows; and Pay-per-Use Audits, allowing SMEs to commission periodic external audits under a standardized framework.

### 4.3.2 Public-Private Partnerships

Regulators, industry groups, and NGOs can collaborate on pooled resources like shared data lakes, open-source fairness toolkits, and joint training for AI risk assessment (Minkkinen et al., 2022). This fosters inclusivity and consistency, smoothing out resource gaps across participants.

### 4.3.3 Enhanced KPI & Measurement Strategies

To prevent reliance on superficial numeric goals, the framework advocates for a mixed-method data collection approach that integrates multiple evaluation strategies. Quantitative metrics include compliance dashboards, adoption rates, and stakeholder diversity indices, providing measurable indicators of engagement and implementation. Qualitative assessments involve interviews, focus groups, and narrative analyses, capturing deeper insights into user trust, conflict resolution, and lived experiences with governance mechanisms. Triangulation (the combination of both quantitative and qualitative streams) ensures a more holistic understanding by not only measuring what was implemented but also assessing how participants experience and perceive the governance process (Leslie, 2019).

### 4.4 Steps Toward Pilot Testing

A pilot applying the framework would follow a five-stage process to ensure structured implementation and evaluation. The Baseline Assessment phase maps existing governance structures, stakeholder roles, and known AI deployments, establishing a reference point. In the Implementation stage, key components such as the Shared Ethical Charter, co-design roundtables, alignment protocols, and oversight boards are introduced or refined. Data Gathering follows, tracking KPI logs (including audit results, user feedback, and cross-jurisdictional compliance) alongside post-deployment interviews to capture stakeholder perspectives. During the Analysis phase, baseline and pilot outcomes are compared to assess improvements in fairness metrics, user satisfaction, and compliance speed. Finally, the Iteration stage adapts the framework based on insights, such as adjusting membership composition or refining the "lite" approach for smaller participants, before re-measuring in subsequent cycles. This cyclical approach fosters continuous learning, ensuring that ethical governance evolves in step with technological and regulatory changes (Omidvar, Kislov & Powell, 2021).

## 5 Discussion

This section reflects on the framework's application, highlighting stakeholder roles, organizational culture, measurement clarity, and global regulatory adaptation.

### 5.1 Conceptual Depth vs. Practical Implementation

The four components bridge principle-based AI ethics and hands-on governance. However, to ensure that these components do not stay at a high level, we introduced vignettes in healthcare, finance, and smart cities. These scenarios illustrate how sandbox testing, representative selection for roundtables, or multi-jurisdiction compliance might unfold. Nonetheless, further refinement through pilot studies or case analyses is warranted to test nuances in different sectors (Cath, 2018; Shams et al., 2023).

Each stakeholder group (developers, regulators, end-users) requires unique operational guidelines. For instance, developers might adopt automated compliance logs that feed into an ecosystem-wide portal, while regulators might prefer formal reporting structures and codes of conduct. End-users (e.g., local communities) typically need simplified, accessible channels for raising concerns or demands. Future expansions can detail how tasks and responsibilities are assigned, specifying the subcommittees or working groups that sustain cross-organizational commitment.

### 5.2 Stakeholder Engagement: The "How" of Co-Design

While co-design is widely advocated, it is rarely operationalized in a structured manner. Our approach emphasizes regular roundtables, recommending a monthly or bimonthly cadence, though frequency should be tailored to the domain, fast-evolving fields like AI-based diagnostics may require more frequent check-ins. Transparent selection of participants is crucial, balancing domain expertise (e.g., medical specialists, financial analysts) with civil society and user voices to prevent corporate capture, where only large vendors shape the conversation (Ansell & Gash, 2008). To address power differentials, discussions should be moderated by a neutral facilitator, often from a third party or NGO, with clear guidelines for conflict escalation and consensus-building to ensure no single interest dominates. Systematic feedback mechanisms, such as an "issue tracker" or "voting mechanism," help record suggestions, highlight adoption rates, and clarify the rationale for rejections. Given the potential conflicts, such as corporate profit motives clashing with fairness demands or regulators advocating for slower deployment, a well-structured co-design process surfaces these tensions early, enabling compromise or strategic redirection of resources (Owen et al., 2012).

### 5.3 Ensuring Measurement Clarity and Avoiding "Tick-Box" Metrics

The risk of superficial compliance, where organizations hit numeric targets without meaningful cultural shifts, looms large (O'Neil, 2017). The framework thus encourages mixed methods. Quantitative metrics (e.g., "Adoption Rate," "Incident Resolution Time") provide essential structure but should be paired with qualitative interviews or participant surveys capturing the quality of stakeholder engagement and the authenticity of leadership support (Leslie, 2019). For instance, an organization may claim a 90% "recommendation adoption" rate from co-design sessions, yet deeper interviews might reveal that critical user concerns were subtly sidelined or that certain community groups remained marginalized.

### 5.4 Organizational Culture, Leadership, and Incentives

Ethical frameworks can fail when organizational or ecosystem incentives prioritize speed, revenue, or cost savings over robust co-design (Floridi, 2019). To counteract this, our analysis highlights the need for leadership accountability, where ethical performance becomes a key part of executive KPIs, ensuring there are consequences for neglecting fairness findings or user complaints. Ethics champions at the middle management level must be empowered to escalate concerns without fear of retaliation, supported by specialized training, direct communication channels, or ethics hotlines. Cultural embedding is also crucial, fostering "learning loops" after each major product cycle so that insights from audits and stakeholder critiques are integrated into future R&D roadmaps (Stahl, Timmermans & Flick, 2017). This requires a multi-level approach: board-level buy-in secures resources and strategic alignment, middle management translates principles into operational workflows, and front-line staff adapt data handling and model updates accordingly (London, 2024).

### 5.5 Resource Constraints and SME Involvement

Scalability is a critical challenge in ethical AI governance. While large multinational corporations can sustain formal oversight boards, repeated audits, and "ethics by design" teams, smaller enterprises often struggle with the cost and complexity of continuous compliance efforts (Shams et al., 2023). To address this, our "lite module" approach reduces overhead through Shared Ethics Boards, where SMEs in a local region or industry consortium pool resources to establish a single body that issues guidelines, reviews compliance, and organizes co-design sessions. Phased implementation allows SMEs to start with simpler checklists or modest annual audits, scaling up as they grow or as the AI system's risk profile increases. Additionally, public funding (such as government incentives or grants for ethical AI pilot programs) can catalyze SME participation and encourage broader adoption of responsible AI practices (Minkkinen et al., 2022). Without such collaborative and scaled approaches, entire ecosystems risk ethical blind spots introduced by smaller vendors or start-ups lacking the capacity for rigorous oversight.

## 5.6 Global Regulatory Adaptation

Adaptive regulatory alignment must account for regional variations in AI governance. While our framework references the EU AI Act (European Commission, 2021), other regulatory frameworks (such as the US's NIST AI Risk Management Framework and China's evolving data laws) necessitate tailored adaptation. Key strategies include mapping, where a matrix comparing major regulatory requirements across jurisdictions (e.g., explainability mandates, prohibited AI practices) ensures that AI solutions meet or exceed local norms. International consortia, such as the Partnership on AI or WEF's Global AI Action Alliance, can help unify guidelines, providing ecosystem participants with a stable compliance baseline. Graduated compliance enables global AI systems to adopt a staged approach, ensuring baseline adherence to the strictest standard (e.g., GDPR for privacy), while allowing for local adaptations that address cultural or legal nuances. While full regulatory harmonization remains unlikely, transparent alignment fosters trust, reduces compliance burdens, and enhances ecosystem consistency (Lin, 2020).

## 5.7 Differentiation from Existing Frameworks

Our framework overlaps with established guidelines, e.g., Leslie, 2019 on process-based governance; IEEE (2019) on ethically aligned design; WEF (2022) on inclusive AI. However, as mentioned in table 1, it further extends them. By uniting ecosystem-level design with co-design and scalable governance structures, we offer a more comprehensive approach that is sensitive to resource disparities and multi-jurisdiction contexts.

Table1. Differentiation from Existing Frameworks

| Dimension | SCOR Framework | Leslie (2019), IEEE (2019), WEF (2022) | Distinctive Feature |
|---|---|---|---|
| Scope | Multi-actor digital ecosystems | Often org. or sector-specific | Emphasizes cross-org synergy & co-design |
| Structure | 4-pillar (Charter, Co-Design, etc.) | Mostly principle-based | Detailed operational steps, vignettes |
| Scale Adaptation | "Lite module" for SMEs | Typically high-level, less SME focus | Offers scaled-down approach & consortia solutions |
| Regulatory Adaptability | Subsection on cross-jurisdiction | Rarely multi-national at length | Stress on "adaptive alignment" & sandboxes |
| KPI Clarity | Mixed method (quant + qual) | Often broad or principle-based | Proposed scoring frameworks, usage logs, etc. |

## 5.8 Future Research Pathways

Future research might evaluate the longitudinal stability of ethics boards (do they maintain authority after multiple product cycles?), explore cross-cultural differences in Charter adoption, or track intangible social outcomes (e.g., user trust, community well-being). Another promising area is pilot testing in high-risk domains, measuring real changes in biases, user complaints, or regulatory compliance cycles (Mäntymäki et al., 2022). By iterating on these insights, the framework can continue evolving, supporting emergent technologies and new forms of cross-border collaboration.

# 6 Conclusion

This paper proposes a four-pillar framework to integrate ethical compliance within AI-driven digital ecosystems. Unlike many guidelines that focus narrowly on single organizations or broad principles, this approach addresses multi-stakeholder networks with varying resource capacities and legal constraints.

Key contributions include an ecosystem perspective that emphasizes synergy among technology providers, regulators, SMEs, and civil society, recognizing that no single actor can ensure ethical outcomes alone; practical protocols that outline structured co-design sessions, auditing practices, and KPI dashboards, offering not just the "why" but the "how" of governance; scalability through a "lite" module strategy for smaller firms with limited budgets or expertise, along with consortia-based solutions for shared ethics boards or sandbox testing; and mixed-method measurement advocating quantitative and qualitative data collection to capture compliance rates, stakeholder perceptions, conflict resolutions, and cultural shifts.

In practice, leaders can embed ethical performance into executive KPIs, finance co-design roundtables, and empower mid-level managers to escalate ethical concerns; regulators can partner with industry in sandboxes, update cross-jurisdiction frameworks regularly, and set transparent guidelines for data usage and fairness checks; and smaller enterprises can collaborate via resource-sharing consortia and adopt scaled-down compliance modules targeting the most critical ethical risks first.

While grounded in design science and scholarly discourse, the framework currently lacks large-scale empirical validation. Pilot implementations (e.g., in finance or healthcare) could document how effectively the components mitigate biases, build user trust, or accelerate regulatory compliance. Cross-cultural or multi-regional studies might reveal how local norms, legal structures, or cultural attitudes influence the Charter's uptake or co-design success. Longitudinal research could assess the sustainability of continuous oversight boards after multiple product iterations or leadership changes (London, 2024).

AI's transformative power requires governance approaches that transcend organizational boundaries, align varied stakeholder incentives, and adapt swiftly to evolving legal contexts. By integrating accountability, diversity/inclusion, and iterative oversight at the ecosystem level, the proposed framework seeks to ensure that AI's benefits are broadly shared and its risks minimized. Serving as both a conceptual scaffold and a practical guide, this four-pillar model paves a path toward responsible innovation across the complex tapestry of modern digital ecosystems.

## References

Adner, R. (2017) 'Ecosystem as structure: An action-based view', Journal of Management, 43(1), pp. 39-58.

Ansell, C. and Gash, A. (2008) 'Collaborative governance in theory and practice', Journal of Public Administration Research and Theory, 18(4), pp. 543-571.

Benjamin, R. (2019) Race After Technology: Abolitionist Tools for the New Jim Code. Polity.

Cath, C. (2018) 'Governing artificial intelligence: ethical, legal and technical opportunities and challenges,' Philosophical Transactions of the Royal Society A, 376(2133), 20180080.

Eubanks, V. (2019) Automating Inequality: How High-Tech Tools Profile, Police, and Punish the Poor. Picador.

European Commission (2021) Proposal for a Regulation Laying Down Harmonised Rules on Artificial Intelligence (Artificial Intelligence Act). Available at: https://eur-lex.europa.eu (Accessed: 20 February 2025).

Floridi, L. (2019) 'Translating principles into practices of digital ethics: Contents, contexts and constraints,' Philosophy & Technology, 32(4), pp. 667-672.

Floridi, L. and Taddeo, M. (2018) 'What is data ethics?,' Philosophy & Technology, 31(4), pp. 123-129.

Gawer, A. (2021) 'Digital platforms and ecosystems: Remarks on the dominant organisational forms of the digital age', Innovation: Organisation & Management, 23(1), pp. 110-124.

Gawer, A. and Cusumano, M. (2014) 'Industry platforms and ecosystem innovation,' Journal of Product Innovation Management, 31(3), pp. 417-433.

Gregor, S. and Hevner, A. R. (2013) 'Positioning and presenting design science research for maximum impact', MIS Quarterly, 37(2), pp. 337-355.

Hevner, A. R., March, S. T., Park, J. and Ram, S. (2004) 'Design science in information systems research,' MIS Quarterly, 28(1), pp. 75-105.

Iansiti, M. and Levien, R. (2004) The Keystone Advantage: What the New Dynamics of Business Ecosystems Mean for Strategy, Innovation, and Sustainability. Boston: Harvard Business School Press.

IEEE (2019) Ethically Aligned Design: A Vision for Prioritising Human Well-being with Autonomous and Intelligent Systems. IEEE.

Jacobides, M. G., Cennamo, C. and Gawer, A. (2018) 'Towards a theory of ecosystems,' Strategic Management Journal, 39(8), pp. 2255-2276.

Jobin, A., Ienca, M. and Vayena, E. (2019) 'The global landscape of AI ethics guidelines,' Nature Machine Intelligence, 1(9), pp. 389-399.

Leslie, D. (2019) Understanding artificial intelligence ethics and safety: A guide for the responsible design and implementation of AI systems in the public sector. The Alan Turing Institute.

Lin, P. (2020) 'Big data ethics and governance: Past, present, and future,' in Makridakis, S. (ed.) Big Data. Wiley, pp. 19-36.

London, A. J. (2024) 'Accountability for responsible AI practices: Ethical responsibilities of senior leadership.' Working paper, Carnegie Mellon University. SSRN: https://ssrn.com/abstract=4736880.

Mäntymäki, M., Minkkinen, M., Birkstedt, T. and Viljanen, M. (2022) 'Putting AI ethics into practice: The hourglass model of organizational AI governance,' arXiv, 2206.00335.

Minkkinen, M., Zimmer, M. P., Eero, S. and Helander, N. (2022) 'Toward systemic AI governance: The roles of policy-makers, data providers, and tech companies.' Government Information Quarterly, in press.

Moore, J. F. (1993) 'Predators and prey: A new ecology of competition,' Harvard Business Review, 71(3), pp. 75-83.

GOV.UK. (2019), New code of conduct for artificial intelligence (AI) systems used by the NHS, https://www.gov.uk/government/news/new-code-of-conduct-for-artificial-intelligence-ai-systems-used-by-the-nhs

O'Neil, C. (2017) Weapons of Math Destruction: How Big Data Increases Inequality and Threatens Democracy. Crown.

OECD (2019) Recommendation of the Council on Artificial Intelligence. OECD/LEGAL/0449.

Omidvar, O., Kislov, R. and Powell, J. (2021) 'The dynamic interplay between knowledge brokering and organisational ethics: A process perspective,' Knowledge Management Research & Practice, 19(3), pp. 276-285.

Owen, R., Macnaghten, P. and Stilgoe, J. (2012) 'Responsible research and innovation: From science in society to science for society, with society,' Science and Public Policy, 39(6), pp. 751-760.

Peffers, K., Tuunanen, T., Rothenberger, M. A. and Chatterjee, S. (2007) 'A design science research methodology for information systems research,' Journal of Management Information Systems, 24(3), pp. 45-77.

Pistilli, G., Muñoz Ferrandis, C., Jernite, Y., & Mitchell, M. (2023). Stronger Together: on the Articulation of Ethical Charters, Legal Tools, and Technical Documentation in ML. In Proceedings of the 2023 ACM Conference on Fairness, Accountability, and Transparency.

Saxena, D., Kahn, Z., Moon E., Chambers, L. M., Jackson, C., Lee M., Eslami, M., Guha, Sh., Erete, Sh., Irani, L., Mulligan, D and Zimmerman, J. (2025). Emerging Practices in Participatory AI Design in Public Sector Innovation, CHI Conference on Human Factors in Computing Systems (CHI EA '25), April 26-May 1, 2025, Yokohama, Japan

Shams, R. A., Zowghi, D. and Bano, M. (2023) 'AI and the quest for diversity and inclusion: A systematic literature review,' AI and Ethics, 5, pp. 411-438.

Smith, C. (2023). AI oversight: Bridging technology and governance, https://www.grantthornton.com/insights/articles/audit/2023/bridging-technology-and-governance

Stahl, B. C., Timmermans, J. and Flick, C. (2017) 'Ethics of emerging information and communication technologies: On the implementation of responsible research and innovation,' Science and Public Policy, 44(3), pp. 369-381.

Stilgoe, J., Owen, R. and Macnaghten, P. (2013) 'Developing a framework for responsible innovation,' Research Policy, 42(9), pp. 1568-1580.

WEF (2022) A Blueprint for Equity and Inclusion in Artificial Intelligence. World Economic Forum White Paper.

Yokoi, T., Wade, M.R. (2023). How organizations navigate AI ethics, International Institute for Management Development, https://www.imd.org/ibyimd/technology/how-organizations-navigate-ai-ethics/

Zowghi, D. and da Rimini, F. (2022) 'Diversity and inclusion in artificial intelligence,' in Handbook of AI Governance and Society. CSIRO Data61.

Appendix 1. SCOR framework details

| Pillar | Details | KPIs | Implementation Vignette |
|---|---|---|---|
| **Shared Ethical Charter (S)** | A binding set of ethical commitments (fairness, accountability, transparency, inclusivity) with baseline requirements (e.g. no discriminatory AI, routine audits, mandatory ethics training, and clear enforcement mechanisms). | • Charter Adoption Rate: % of organizations formally endorsing and integrating the charter.<br>• Ethics Training & Awareness: % of staff completing annual ethics training with high assessment scores.<br>• Ethical Decision Review Coverage: % of high-impact AI projects vetted pre-deployment.<br>• Charter Compliance Incidents Resolved: % of reported breaches remediated. | In healthcare, a national service (e.g. the NHS) partners with tech firms to develop an AI code of conduct. Oversight committees review patient care systems for bias, with public logs and mandatory staff training ensuring accountability. |
| **Co-Design and Stakeholder Engagement Mechanisms (C)** | Regular, structured co-design sessions with diverse stakeholders (domain experts, underrepresented groups, NGOs, regulators) and rotating representation. Documentation and neutral facilitation ensure that stakeholder input is meaningfully integrated into design decisions. | • Stakeholder Inclusion in AI Projects: % of projects involving external consultations.<br>• Diversity of Stakeholder Representation: Index score assessing demographic and expertise balance.<br>• Stakeholder Feedback Implementation Rate: % of co-design recommendations adopted.<br>• Stakeholder Impact Assessment Completion: % of high-risk AI projects undergoing formal impact reviews.<br>• Stakeholder Trust/Satisfaction Score: Aggregated survey feedback on inclusiveness and transparency. | In a cross-border fintech alliance, bimonthly roundtables bring together consumer advocates, data scientists, and regulators. Public meeting summaries and formal assessments ensure that input shapes credit-risk models. |
| **Oversight and Learning (O)** | Continuous oversight through periodic (scheduled/random) audits, an internal/external governance board, and a shared knowledge repository for best practices and incident logs. | • Internal AI Governance Committee Presence: % of organizations with dedicated AI ethics committees.<br>• Independent Audit Coverage: % of high-risk systems externally audited annually.<br>• Audit Trail & Documentation Quality: % of projects with complete audit logs.<br>• Oversight Recommendations Implementation: % of action items executed post-audit.<br>• Incident Reporting & Remediation Rate: Ratio of reported incidents resolved. | A global software company establishes a dual oversight model (internal committee plus an external advisory board) to review high-risk projects continuously, ensuring prompt corrective actions and shared learning via a centralized repository. |
| **Regulatory Alignment (R)** | An adaptive framework that anticipates evolving AI regulations through regulatory sandboxes, horizon scanning committees, and cross-border interoperability, ensuring high-risk AI meets strict legal standards. | • High-Risk AI Compliance Rate: % of high-risk systems meeting all regulatory requirements.<br>• Regulatory Audit/Certification Pass Rate: % of systems passing external audits or achieving certifications.<br>• Timely Regulatory Update Implementation: Median time to update processes after regulatory changes.<br>• Mandatory Impact Assessment Completion: % of required impact assessments completed.<br>• Regulatory Incident Rate: Annual count of legal violations or fines. | A multinational bank launching an AI credit scoring system collaborates with regulators across regions to ensure compliance. Regular cross-jurisdictional audits and rapid process updates build trust with both regulators and customers. |